\documentclass{aa}  

\usepackage{graphicx}
\usepackage{multicol}
\usepackage[toc,page]{appendix}

\usepackage{txfonts}
\usepackage{hyperref}
\begin{document}

     \title{Turnaround density evolution encodes cosmology in simulations}

   \author{Giorgos Korkidis
          \inst{1}\fnmsep\inst{2}\fnmsep
          \thanks{E-mail: gkorkidis@physics.uoc.gr},
          \
          Vasiliki Pavlidou \inst{1}\fnmsep\inst{2},
          \and
          Konstantinos Tassis \inst{1}\fnmsep\inst{2}
          }
    
    \authorrunning{Korkidis, Pavlidou \&  Tassis}

   \institute{Department of Physics and Institute for Theoretical and Computational Physics, University of Crete, GR-70013 Heraklio, Greece
         \and
             Institute of Astrophysics, Foundation for Research and Technology – Hellas, Vassilika Vouton, GR-70013 Heraklio, Greece
             }

   \date{}
 
  \abstract
   {The mean matter density within the turnaround radius, which is the boundary that separates a nonexpanding structure from the Hubble flow, was recently proposed as a novel cosmological probe. According to the spherical collapse model, the  evolution with cosmic time of this turnaround density, $\rho_{ta}(z)$, can be used to determine both $\Omega_m$ and $\Omega_\Lambda$, independently of any other currently used probe. The properties of  $\rho_{ta}$ predicted by the spherical collapse model (universality for clusters of any mass, value) were also shown to persist in the presence of full three-dimensional effects in $\Lambda$CDM N-body cosmological simulations when considering galaxy clusters at the present time, $z=0$. However, a small offset was discovered between the spherical-collapse prediction of the value of $\rho_{ta}$ at $z=0$ and its value measured in simulations.}
   {In this letter, we explore whether this offset evolves with cosmic time; whether it differs in different cosmologies; whether its origin can be confidently identified; and whether it can be corrected. Specifically, we aim to examine whether the evolution of $\rho_{ta}$ can be used to distinguish between simulated universes with and without a cosmological constant.}
   {We used N-body simulations with different cosmological parameters to trace the evolution of the turnaround density $\rho_{ta}$ with cosmic time for the largest dark matter halos in the simulated boxes. To this end, we analyzed snapshots of these simulations at various redshifts, and we used radial velocity profiles to identify the turnaround radius within which we measured $\rho_{ta}.$}
   {We found an offset between the prediction of the spherical collapse model for $\rho_{ta}$ and its measured value from simulations. The offset evolves slightly with redshift. This offset correlates strongly with  the deviation from spherical symmetry of the dark matter halo distribution inside and outside of the turnaround radius. We used an appropriate metric to quantify deviations in the environment of a structure from spherical symmetry. We found that using this metric, we can construct a sphericity-selected sample of halos for which the offset of $\rho_{ta}$ from the spherical collapse prediction is zero, independently of redshift and cosmology.}
   {We found that a sphericity-selected halo sample allows us to recover the simulated cosmology, and we conclude that the turnaround density evolution indeed encodes the cosmology in N-body simulations. }


   \maketitle
%

\section{Introduction} \label{section 1}
The question of assigning meaningful boundaries to large-scale structures has traditionally been addressed via two distinct paths. The first path involves overdensity criteria: The boundary of the structure is defined as the scale at which the average matter density becomes some given multiple of the mean matter density, or of the critical density, of the Universe at the time of observation. Overdensity boundaries are motivated by the spherical collapse model (e.g., \citealp{GG1972,Gunn1977,Lahav1991}).
The second path involves kinematic or dynamic criteria, motivated by the study of infall onto existing collapsed structures (e.g., \citealp{FG1984,Bertschinger1985}). Most recently, this approach motivated the introduction of the splashback radius
(\citealp{DK2014, Adhikari2014, More2015}) as a  boundary for structures that are still accreting matter from their environment. 

A hybrid of the two types of boundaries is the turnaround radius, which is a kinematically motivated boundary (the scale that separates a nonexpanding structure from the Hubble flow) that according to spherical collapse, also constitutes a scale of constant overdensity for collapsed structures of all masses at a given redshift.  In recent years, the turnaround radius has gained considerable attention as a scale on which cosmological models can be tested (e.g., \citealp{Pavlidou_Tomaras2014, TanoglidisPavlidouTomaras2015, TanoglidisPavlidouTomaras2016, tests8, tests7, tests6, tests5,  PavlidouEtal2020, tests4, tests3, tests2, Korkidis_etal, tests1}).

 \cite{PavlidouEtal2020} showed by employing the spherical-collapse model that the matter density within the turnaround scale (the turnaround density, $\rm \rho_{ta}$) could be used as a cosmology-probing observable, with a number of attractive properties: For a given redshift, the turnaround density is universal and thus insensitive to halo size, selection biases, and sample completeness issues; its present-day value almost exclusively probes the matter density; and its evolution with redshift is sensitive to the dark energy content of the Universe. 

At the same time, the assumption of spherical symmetry implicitly used in \cite{PavlidouEtal2020} contradicts the highly nonspherical nature of realistic cosmological structures, raising serious concerns about the practical utility of $\rho_{\rm ta}$ as a quantitative cosmological probe. However,  \cite{Korkidis_etal} showed with the aid of $\rm \Lambda CDM$ cosmological N-body simulations that at the present time (z=0), a dynamically meaningful turnaround radius can be measured kinematically from radially averaged velocity profiles for realistic cosmological structures. The average matter density within that scale was also shown to have a value that is very close to the value predicted by the spherical-collapse model. These results were consistent with earlier studies of simulated structures on similar scales (e.g., \citealp{Busha_Wechsler_2005,Cupani2008}).

In this paper, extending the work of \cite{Korkidis_etal}, we aim to show that this agreement between the turnaround dynamics in realistic 3D structures and the spherical-collapse model persists for higher redshifts and for cosmological simulations with underlying cosmologies different from the concordance $\rm \Lambda CDM$. In \citet{Korkidis_etal}, the kinematically measured $\rm \rho_{ta}$ had a systematic offset with respect to the prediction of the spherical collapse model. This offset would make the turnaround density problematic as a cosmological observable. In principle, this offset might be calibrated away using simulations before comparing with observations of clusters. However, this would be increasingly complicated if the offset depended on redshift and/or cosmology, and it would be impractical if these dependences were such that they would cause $\rho_{\rm ta}(z)$ curves of different cosmologies to appear similar. The ideal situation would be to identify the origin of the offset and, if possible, eliminate it by applying appropriate quality cuts to the cluster sample used to constrain the cosmological parameters. Exploring these possibilities is a second objective of this work. 

This paper is organized as follows: In \S \ref{sims} we describe the cosmological N-body simulations that we employed, the halo sample we used, and the method for calculating the turnaround density. In \S \ref{section 3} we present our results. Specifically we show how $\rm \rho_{ta}$ evolves with cosmic time for different cosmologies; how the offset of $\rm \rho_{ta}(z)$ from spherical-collapse predictions changes with redshift and cosmology; that the offset, as would intuitively be expected, is dependent on the distribution of neighboring halos in the vicinity of the turnaround radius of a structure, and that we can use this fact to design appropriate cluster selection criteria for the elimination of the offset. We summarize and discuss our findings in \S 4.

\section{Simulated data and methods}\label{sims}

\begin{table*}[htb!]
\begin{center}
\caption{Cosmological parameters used in each simulation}
\label{table 1}
\begin{tabular}{c c c c c c}
\hline
Simulation & $\Omega_{m,0}$ & $\Omega_{{\Lambda,0}}$ & $h_{100}$ & $\sigma_8$ & $\rm{\epsilon\left [ h^{-1} kpc  \right ]}$ \\ \hline
Virgo $\rm \Lambda CDM$  & $0.3$ & $0.7$ & $0.7$ & $0.90$ &  $25$ \\
Virgo $\rm OCDM$ & $0.3$  & $0$  & $0.7$ & $0.85$ & $30$ \\
Virgo $\rm SCDM$ & $1$  & $0$  & $0.5$ & $0.51$ & $36$ \\
MDPL2 & $0.307$ & $0.693$ & $0.6770$ & $0.81$ & $2$ \\ \hline
\end{tabular}
\end{center}
\end{table*}

\subsection{\textbf{Cosmological N-body simulations}}

In this section, we describe the cosmological simulations that we used in our analysis along with the methods that we employed in order to measure the matter density at turnaround $\rm \rho_{ta}$. One of the main characteristics of the turnaround density in the spherical collapse model is that its evolution with cosmic time is sensitive to the presence of a cosmological constant. Thus, one of our objectives is to show that in N-body cosmological simulations, the kinematically measured $\rm \rho_{ta}$ is also sensitive to the input cosmological parameters. To this effect, we analyze simulated halos from simulations assuming three different cosmologies: two flat   models with $\Omega_m \sim 0.3$ and $\Omega_\Lambda \sim 0.7$ (referred to as $\rm \Lambda CDM$), an open-matter-only model with $\Omega_m \sim 0.3$ and no cosmological constant ($\rm OCDM$), and a flat-matter-only model, with $\Omega_m = 1$ and no cosmological constant ($\rm SCDM$).


One of our $\rm \Lambda CDM$ dark-matter-only boxes was taken from the MupltiDark Planck simulations (MDPL) \citep{MDPL_simulations}, and it spans $\rm L = 1000 \ h^{-1} \ Mpc$ on a side. This particular run followed the evolution of $3840^3$ particles with individual masses of $\rm M_p = 1.51 \times 10^{9} \ h^{-1}  \ M_{\odot}$ (MDPL2) and a cosmology consistent with Plank measurements (\cite{Plank2016}; see \ref{table 1}).

The second $\rm \Lambda CDM$ box as well as the  $\rm OCDM$ and $\rm SCDM$ simulations we analyzed were taken from the Virgo consortium\footnote{https://wwwmpa.mpa-garching.mpg.de/Virgo/virgoproject.html} suit of cosmological simulations. In particular, we analyzed their intermediate-sized N-body runs, which follow the evolution of $256^3$ particles with mass $\rm M_p = 6.86/22.7 \times 10^{10} \ h^{-1}  \ M_{\odot} \ (\Lambda CDM, OCDM/SCDM)$ in a box of $\rm L = 239.5 \ h^{-1} \ Mpc$ on a side.

Our choice to analyze older cosmological runs was driven by efficiency considerations (using available data rather than unnecessarily performing N-body runs from scratch) in combination with the fact that as recent cosmological data disfavor cosmologies other than $\rm \Lambda CDM$, virtually all currently produced publicly available datasets are small variations of $\Lambda CDM$, as effort has shifted toward running cosmological boxes with increasingly complex baryonic physics, or nonstandard dark matter models.

While no longer meeting the needs of most current cosmology projects in terms of their box size and resolution, the Virgo simulations were at the frontier of cosmological research at the time they were made publicly available, and they have been  widely used and tested. We therefore chose to work with this very thoroughly explored set of cosmological runs because they are already publicly available and include models that are sufficiently different from the currently preferred concordance flat $\rm \Lambda CDM$ so that the sensitivity of $\rm \rho_{ta}(z)$ to cosmological parameters can be tested effectively.  
We additionally emphasize that the implementation of gravity in cosmological simulations has not changed since the Virgo runs, so  given that our box size and resolution requirements are satisfied, the Virgo runs are adequate for our purposes. 

Regarding halo identification, in the case of the MDPL2 box, the halo catalog was produced using the Rockstar algorithm \citep{Rockstar}. For the Virgo simulations, we used the friends-of-friends algorithm (FoF; \cite{FoF}) using nbodykit \footnote{Nboykit is a massively parallel large-scale structure toolkit in python. For more details: \\ https://nbodykit.readthedocs.io/en/0.1.11/index.html} , and then we implemented a spherical overdensity (SO) criterion to calculate $\rm M_{200}$ masses for the identified halos.  snapshots that we used from both suits of simulations were in the range $\rm 0 \leq z \leq 1$.\footnote{This redshift range is more than adequate: in \cite{PavlidouEtal2020}, we explicitly showed that in the context of the spherical-collapse model, by measuring $\rm \rho_{ta}$ and its evolution with cosmic time, we can distinguish between different cosmological models by a redshift $\rm z = 0.3$ if observations of thousands of clusters are available at that z and they do not suffer from any offset.}

\subsection{\textbf{Turnaround density calculation, halo sampling, and substructure elimination}}

To calculate the turnaround density in simulations, the turnaround radius (the cluster-centric distance at which dark matter particles (on average) join the Hubble flow) and the turnaround mass (the total mass enclosed by the turnaround radius) have to be measured. In \citet{Korkidis_etal} we showed that the turnaround radius for group- and cluster-sized halos is well defined as the radial shell for which the average radial velocity of dark matter particles, $\langle V_{r}\rangle$, crosses zero for the first time as the distance from the center of the cluster decreases. This shell also lies in a region in which the velocity dispersion $\left< V_{r}^2\right>$ is minimum. This feature can also be identified in particle phase-space diagrams of cluster-sized halos (e.g., \citealp{Cuesta2008,VogelsbergerWhite2011} ).

At each redshift snapshot of the Virgo simulations, we analyzed the 1000 most massive halos. 

At $\rm z=0,$ this corresponded to halos with masses in the interval $\rm{ 10^{14} M_\odot\leq M_{200} \leq 10^{15} M_\odot}$ . Most of them lie at the lower limit. For each of these halos, we calculated the turnaround radius, $\rm R_{ta}$, and then proceeded to identify substructures within this scale.  

For this task, we followed the analysis of \cite{Korkidis_etal}, where for each of our halos, we identified all neighboring halos within its $\rm R_{ta}$ and labeled as "structures" those with the largest $\rm M_{200}$, discarding the remaining "substructure" halos from further analysis. Through this cleaning procedure, $3-7 \%$ of the structures at each redshift were discarded as substructures for all of our simulated boxes.

A similar strategy was also followed for the MDPL2 simulation, but because in this case the simulated box was considerably larger, we followed a different  approach to select the halo sample. In particular, we considered two samples: one sample with the 3000 most massive halos of the box ($\rm{ 10^{14} M_\odot\leq M_{200} \leq 6 \times 10^{15} M_\odot}$), and a second sample for which we randomly selected 780 halos from logarithmically spaced mass bins, with $\rm M_{200}\geq M_{\rm min}$. The value for $M_{\rm min}$ at each redshift snapshot scaled proportionately to the mass of the largest halo at that time, and was equal to $6 \times 10^{13} M_\odot $ for $z=0$. 

The reasoning for choosing a random sample extending to low masses was twofold. On the one hand, we wished to examine a large enough dynamical range in mass (at least two orders of magnitude) so as to be able to identify any mass-driven effects or correlations. On the other hand, because of the enormous size of the halo catalog, we had to use some trimming-down algorithm for the halo catalogue to keep our analysis computationally tractable.

\section{Results} \label{section 3}
We used the simulated clusters and methods described in the previous section to study how the turnaround density $\rm \rho_{ta}$ evolves with cosmic time, compare it with the predictions of the spherical collapse model, and trace the source of deviations between the two.

\subsection{\textbf{Offset of the turnaround density in simulations from the spherical collapse prediction evolves  with cosmic time}} \label{subsection 3.1}

In \citet{Korkidis_etal}, we showed that the turnaround radius $R_{\rm ta}$ is well correlated with the turnaround mass $M_{\rm ta}$, with a scaling very close to $R_{\rm ta}^3 \sim M_{\rm ta}$, so that a meaningful characteristic turnaround density exists for $z=0$ in a $\Lambda$CDM cosmology. We have confirmed that this remains true across redshifts and cosmologies (see Appendix~\ref{appendix:a}). In this section, we compare the value of this characteristic (average) turnaround density with spherical-collapse predictions.

\begin{figure}[htb!]
    \includegraphics[width=1.07\columnwidth,clip]{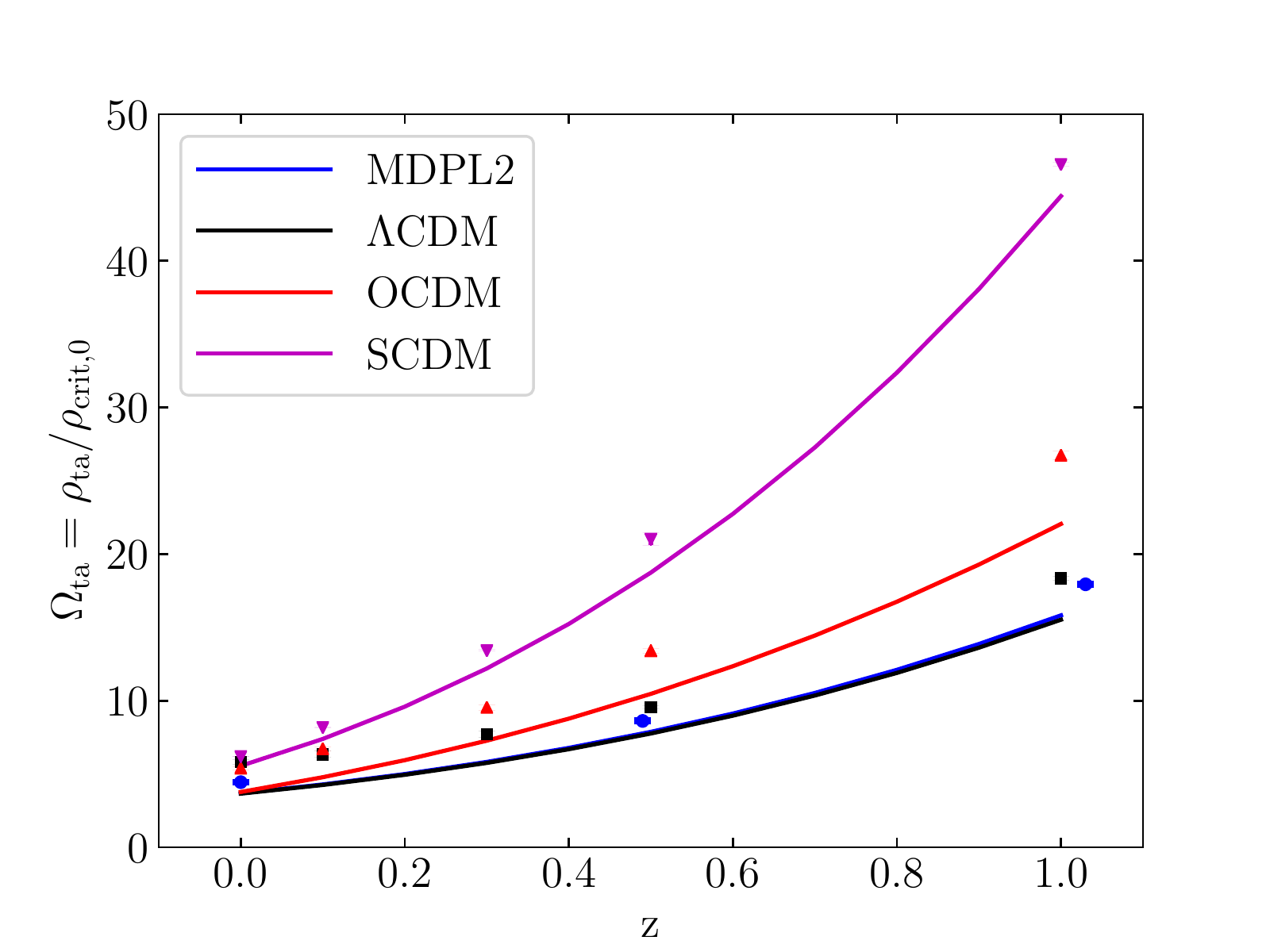}
    \includegraphics[width=1.07\columnwidth,clip]{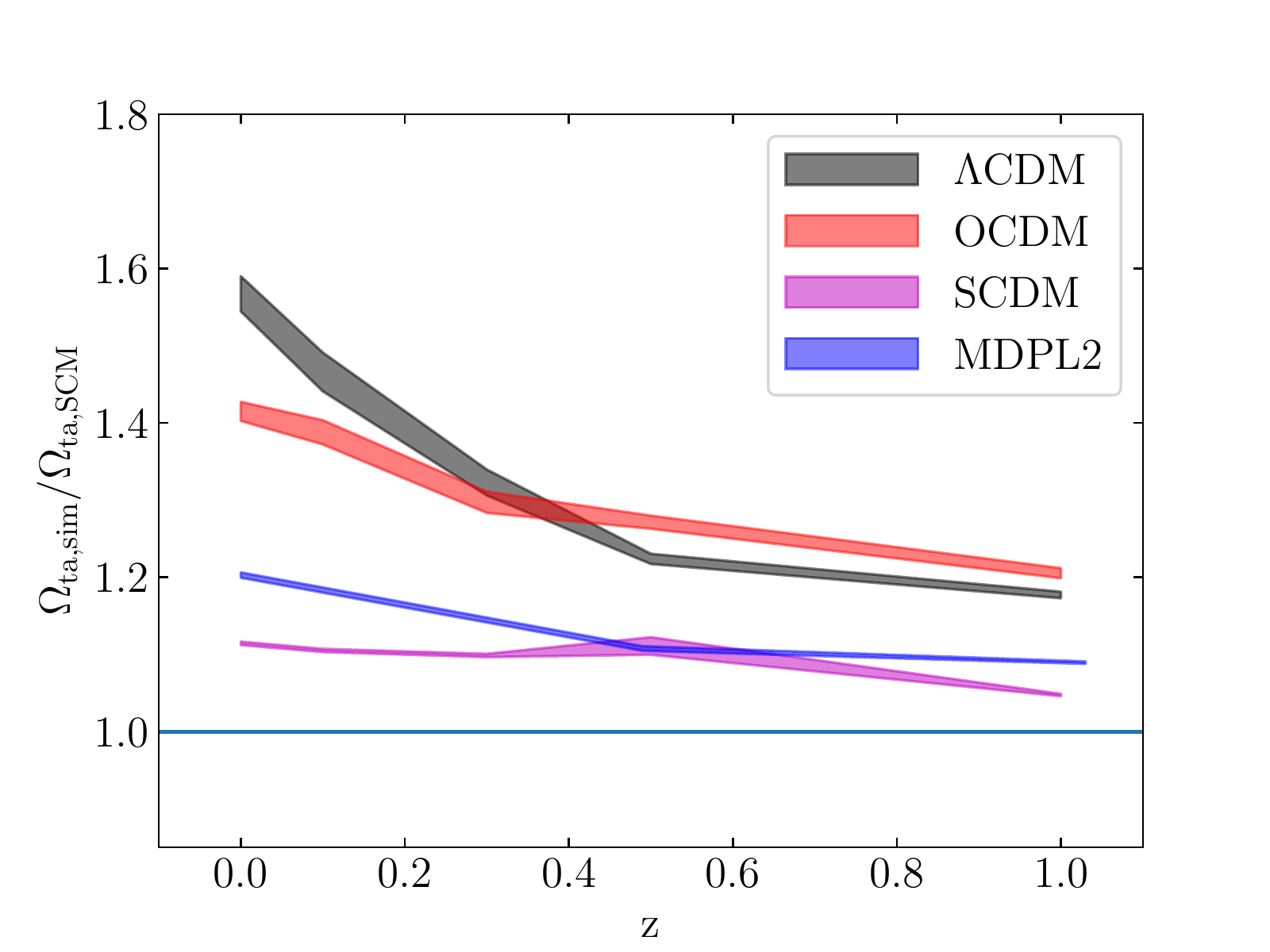}
\caption{Evolution of the turnaround density with redshift. Upper panel: Data points: $\rm \Omega_{ta}$ vs. redshift for the three Virgo cosmologies [$\rm \Lambda CDM$ (black points), $\rm OCDM$ (red points), and $\rm SCDM$ (magenta points)] and for the MDPL2 $\rm \Lambda CDM$ run. Each point corresponds to the mean value of $\rm \Omega_{ta}$ in the halo sample at that $z$, and the error bars (in most cases smaller than the symbol size) correspond to the standard error on the mean. Solid lines: Spherical-collapse prediction for $\rm \Omega_{ta}$ vs. $z$ for the corresponding cosmology. Lower panel: Ratio of $\rm \Omega_{ta}$ in simulations over the spherical-collapse prediction  as a function of z for the Virgo and MDPL2 simulations. There is a small, evolving offset between the kinematically measured turnaround density in simulated halos and the spherical-collapse prediction. }
\label{Fig. 1.}
\end{figure}

The data points in the upper panel of Fig.~\ref{Fig. 1.} depict the evolution of the turnaround density with redshift for the three Virgo cosmologies and for MDPL2. The density was normalized with respect to the present-day value of the critical density $\rm \rho_{crit,0}$ in each cosmology. 
The solid lines represent the evolution of $\rm \Omega_{ta}$ , with $z$ predicted from the spherical-collapse model  \citep{PavlidouEtal2020} in each cosmology. 

From the figure, two things are apparent for all three cosmologies: (a) As early as $\rm z=0.3-0.5$, the $\Omega_{ta}$ values of different cosmologies start to diverge.  (b) There is a systematic offset between the simulation points and the spherical-collapse lines that  evolves (decreases) weakly with cosmic time.

The second point is made more clear in the lower panel of Fig.~\ref{Fig. 1.}, where we plot the ratio of $\rm \Omega_{ta}$ from simulations and its predicted value from the spherical collapse model. Black, red, and magenta shaded regions represent the mean values of the simulations plus or minus one standard error 
for the three Virgo runs, while blue corresponds to the results from the MDPL2 simulation. Even accounting for halo-to-halo spread, it is clear that for all simulation suits, the offset decreases with increasing redshift. For the Virgo simulations, the offsets for $\rm \Lambda CDM$ and $\rm OCDM$ are very similar up to $\rm z=0.3,$ while the $\rm SCDM$ offset is much lower at all redshifts and even becomes negligible for high $\rm z$. The MDPL2 simulation appears to have a much smaller offset than its Virgo $\rm \Lambda CDM$ counterpart, although we should highlight that the two samples are very different in terms of the number of halos analyzed in all mass ranges. In any case, it would appear that the origin of the offset is not the cosmology itself (in which case the MDPL2 and Virgo $\rm \Lambda CDM$ data sets would feature very similar offsets), but some property of the halos in the sample. 

Understanding the source of this offset is of paramount significance for any future attempt to measure $\rm \rho_{ta}$ and its evolution with cosmic time. In the remainder of this section, we therefore attempt to identify the source of the offset and its evolution.

\subsection{\textbf{Offset correlates with local  (a)sphericity}} \label{subsection 3.2}
 
 In \cite{Korkidis_etal}, we demonstrated that in the case of a $\rm \Lambda CDM$ simulated Universe, the absolute fractional deviation of the value of $\rm R_{ta}$ from its predicted value from the spherical collapse model at $\rm z=0$ was weakly correlated with the deviation of a structure from spherical symmetry and with the presence of massive neighbors outside of the turnaround radius. Here we complement this analysis by investigating the impact of the environment on the turnaround density offset more thoroughly. In particular, we consider the effect of halos adjacent to the primary cluster both inside and outside of the turnaround radius.
 
By definition, the turnaround density is very sensitive to the location of the turnaround radius $\rm R_{ta}$ , and to a lesser extent, to the enclosed mass $\rm M_{ta}$: Because dark matter halos are highly concentrated, changes in $\rm R_{ta}$ will not affect the enclosed mass dramatically, but will very strongly affect the turnaround volume. Hence, if we were to search for the origin of deviations of $\rm \rho_{ta}$ from any reference density (e.g., the $\rm \rho_{ta}$ predicted by the spherical-collapse model), we would mainly have to consider the factors that affect the location of $\rm R_{ta}$. These factors should be gravitational in nature, and more specifically, they must be related to the distribution of matter in the vicinity of the turnaround radius. 

A metric that pertains to how matter is distributed is the one we used in \cite{Korkidis_etal} to describe the extent to which matter deviates from spherical symmetry. We calculated this metric as follows: 
\begin{equation}
     \alpha_{3D} = \frac{I_{k,min}}{I_{k,max}}
,\end{equation}
    where $I_k$  with  $k=1,2,3$ the principal moments of inertia. A value of 1 accordingly corresponds to a sphere, whereas a value of 0 corresponds to prolate or oblate objects of infinitesimal thickness.

\begin{figure*}[htb!]
\begin{multicols}{2}
    \includegraphics[width=\linewidth]{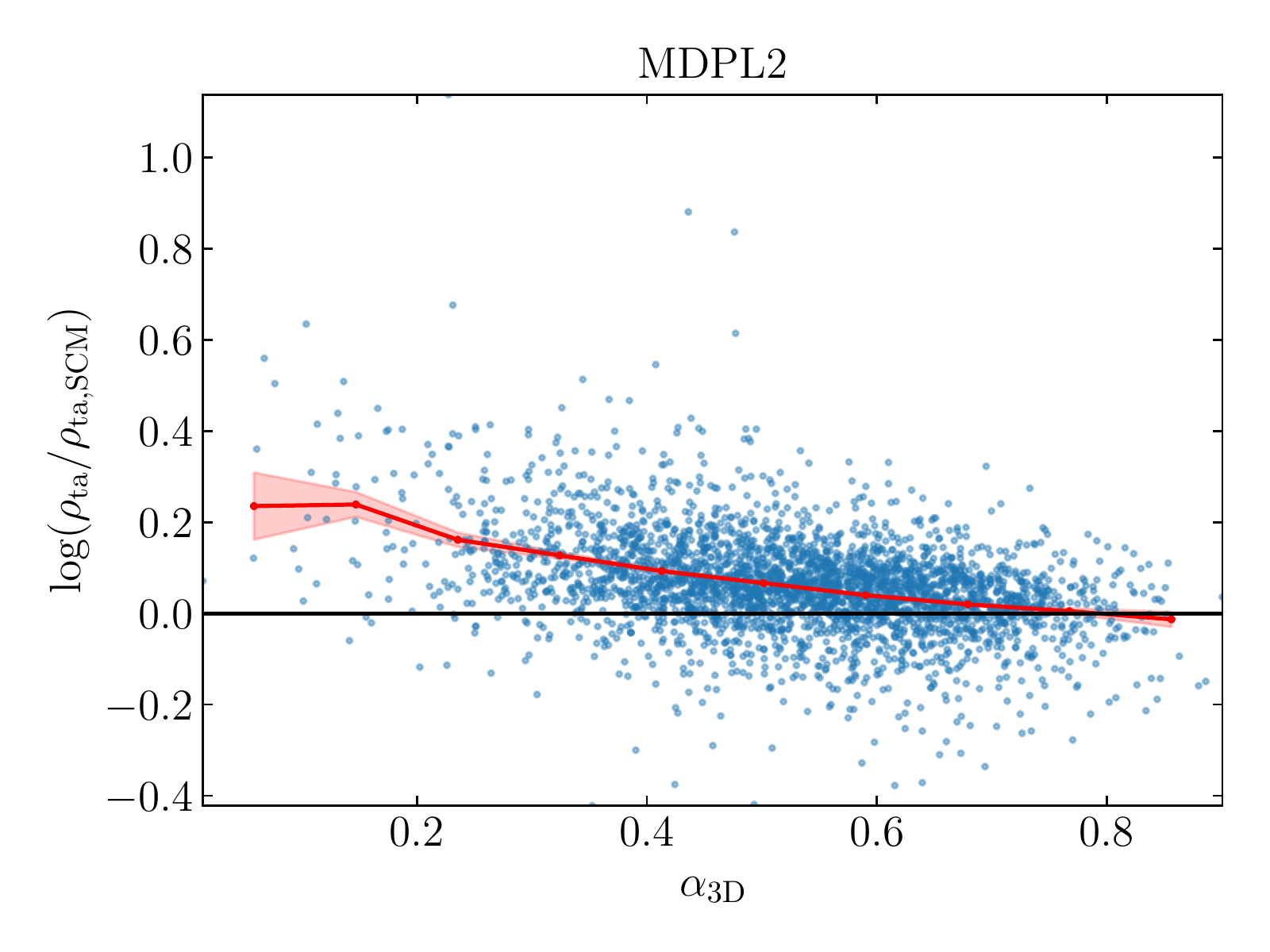}\par 
    \includegraphics[width=\linewidth]{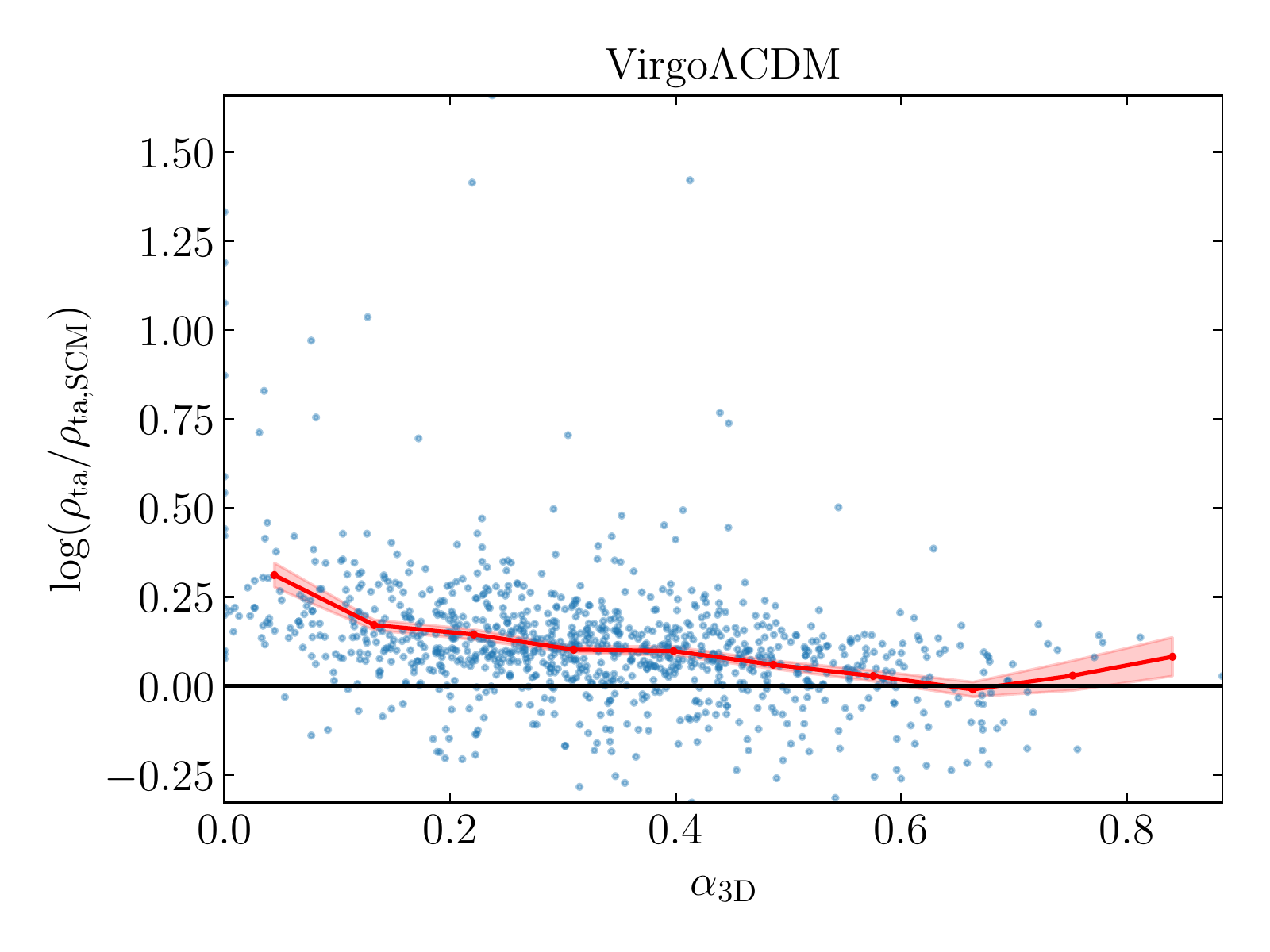}\par 
    \end{multicols}
\begin{multicols}{2}
    \includegraphics[width=\linewidth]{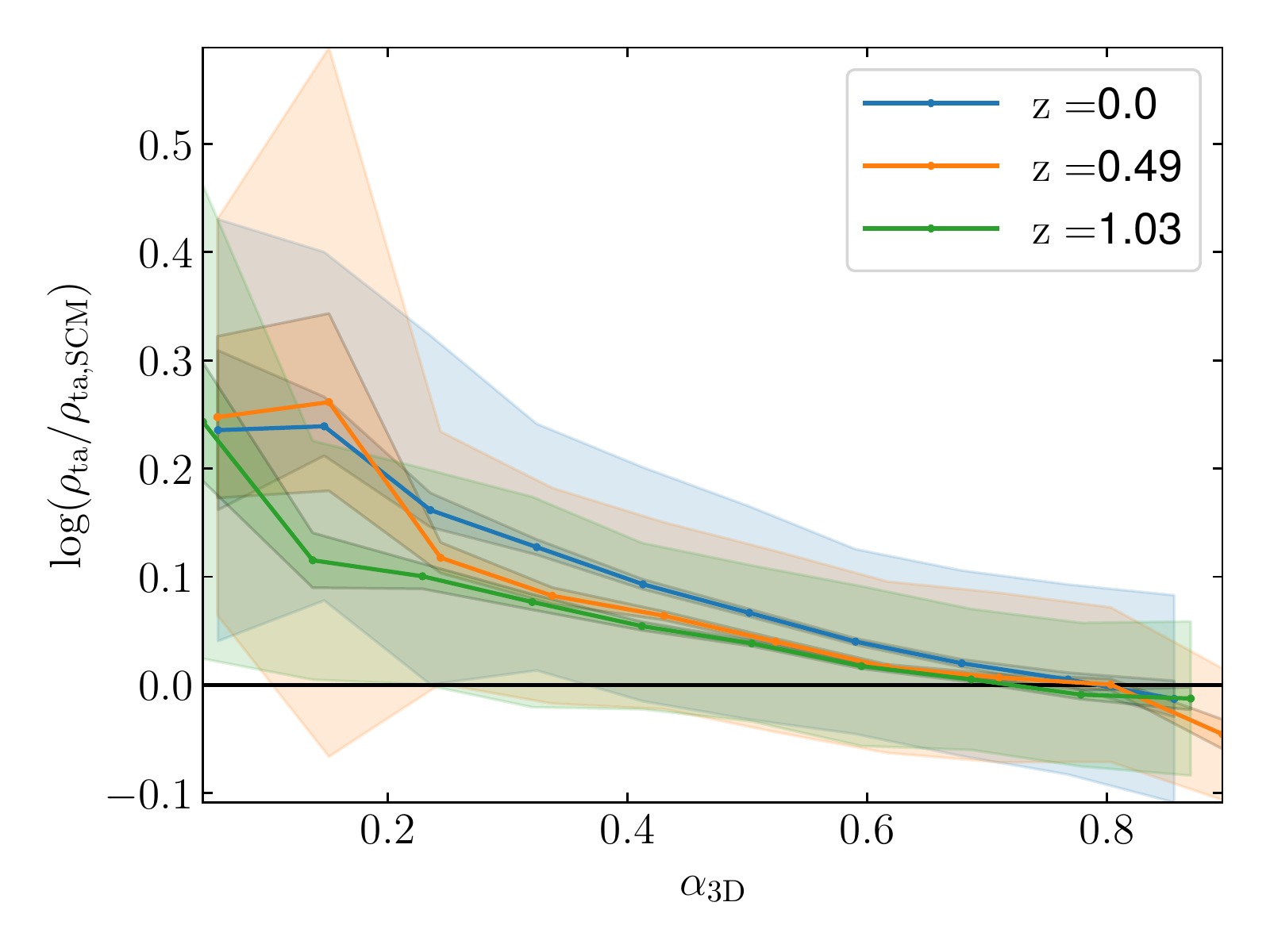}\par
    \includegraphics[width=\linewidth]{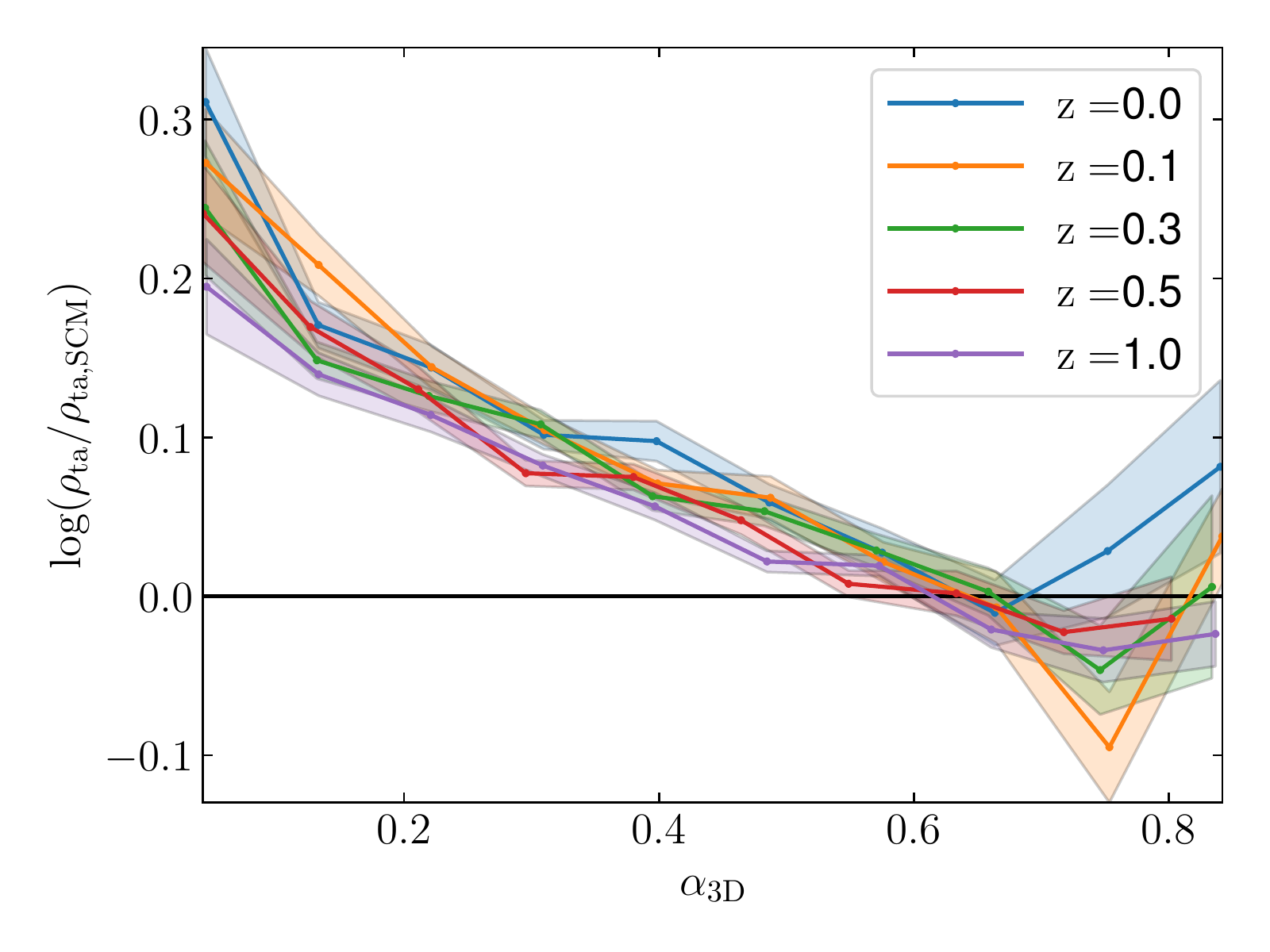}\par
\end{multicols}
\caption{Correlation between the turnaround density offset of a structure and the deviations of its neighbours from spherical symmetry. Upper panel: Blue points show the turnaround density offset $\rm \rho_{ta}/ \rho_{ta,SCM}$ as a function of the sphericity metric $\alpha_{3D}$ (see text) for MDPL2 and $\rm Virgo \Lambda CDM$  halos at a redshift $\rm z=0$. The two quantities are negatively correlated, as is confirmed by a Spearman test: very significant (very low p-value) moderate correlation, with a correlation coefficient of -0.42 and -0.39 for the two simulations, respectively. The red line shows the mean value of the offset in bins of $\alpha_{3D}$. The error in the y-axis depicts the standard error of the mean. The lower panel shows the same as the red points in the upper panel. Different colors represent different redshifts in MDPL2 and $\rm Virgo \Lambda CDM$. In the case of MDPL2 (left panel), we also included a low-tone shaded region showing the $\mathbf{\rm 1 \sigma}$ spread of points around the mean.}
\label{Fig. 2.}
\end{figure*}

For this work, we calculated the sphericity  $\alpha_{3D}$ of structures surrounding the primary clusters, using the locations and masses of all objects in the halo catalog with mass $\rm \geq 10^{12} M_{\odot}$ and with a distance from the primary clusters in the interval $\rm R_{200} < R \leq 1.5 \ R_{ta}$. The results from this analysis are shown in the upper panel of Fig.~\ref{Fig. 2.}, where we plot the ratio of the turnaround density over the spherical collapse prediction as a function of $\alpha_{3D}$ for MDPL2 and $\rm Virgo \Lambda CDM$ halos at a redshift $\rm z=0$ with blue dots. 

Even by inspection of this figure alone, it is clear that  these two quantities are negatively correlated, as would intuitively be expected: the closer the dark matter halo distribution around a halo is to spherical symmetry, the smaller the offset of the turnaround density from the spherical collapse expectation. 
In order to make the trend even more apparent, we overplot in red the mean value of the offset for bins of the sphericity metric $\alpha_{3D}$. The offset plateaus at zero (the ratio of $\rho_{\rm ta}$ over the spherical collapse prediction plateaus at 1) for $\alpha_{3D} \gtrsim 0.6.$

This correlation is robust with cosmic time, as is shown in the lower panel of Fig.~\ref{Fig. 2.}, where we plot again the mean value of the offset in $\alpha_{3D}$ bins for MDPL2 halos of various redshifts up to a redshift of $\rm z\sim 1$. Again, for $\alpha_{3D} \gtrsim 0.6,$ the offset tends to scatter uniformly around zero. This trend holds for all cosmologies tested in this work. 

\subsection{\textbf{Offset evolution is eliminated by sphericity cuts}} \label{subsection 3.3}

We have established that the deviation of the $\rho_{ta}$ of a structure from the prediction of the SCM is strongly correlated with the deviation of the distribution of its neighbors from spherical symmetry and that, importantly, this correlation remains robust up to a redshift of one. Hence, we hypothesize that we can use $\rm \alpha_{3D}$ as a selection criterion to eliminate the offset that plagues the measurement of the turnaround density. That is, if we were to impose a cut in our analysis and examine halos with $\rm \alpha_{3D} \geq 0.5,$ we might be able to cause the offset to disappear. 

\begin{figure}[htb!]
    \includegraphics[width=1.07\columnwidth,clip]{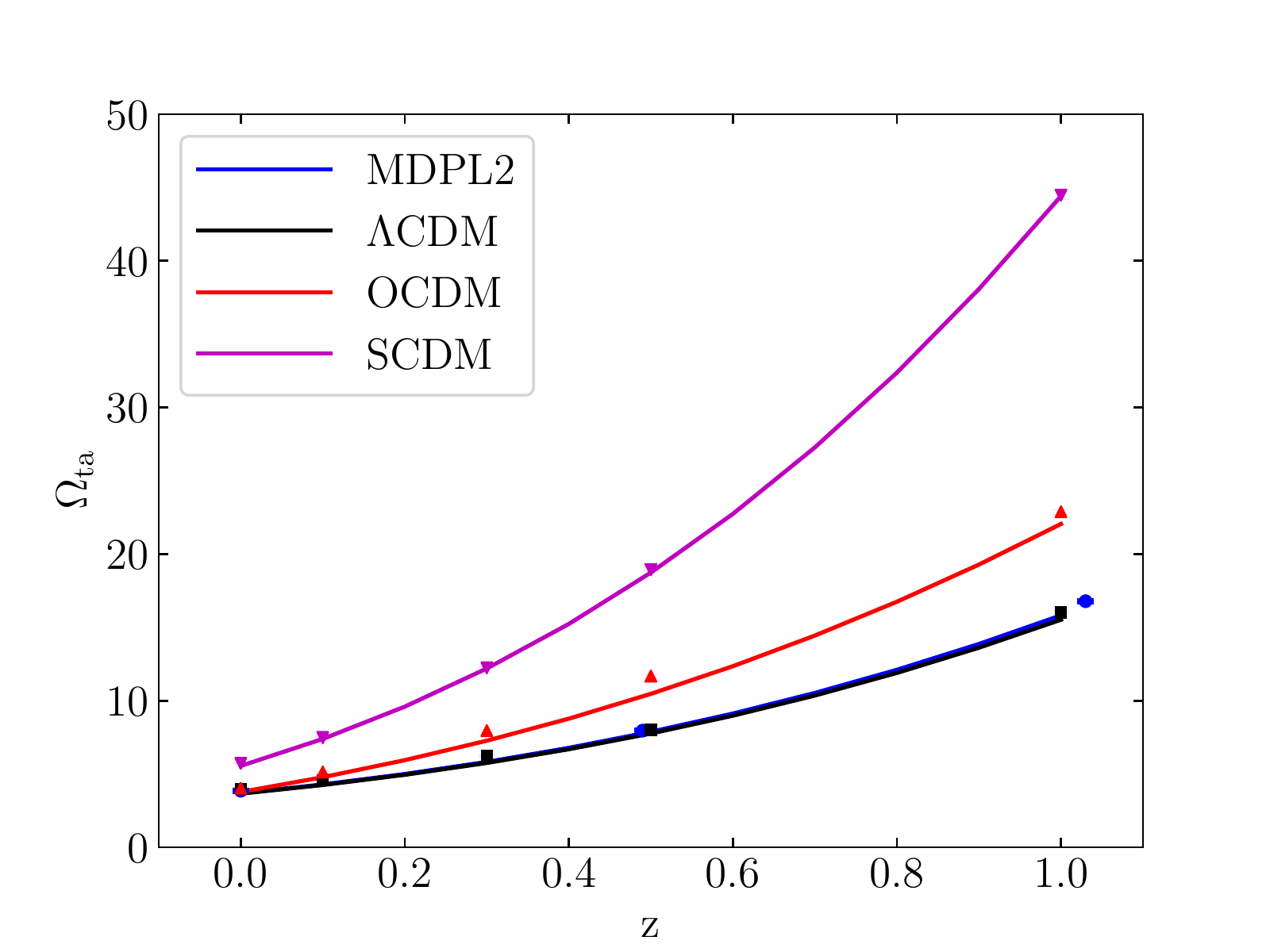}
\caption{As in the upper panel of Fig.~\ref{Fig. 1.}, we plot the evolution $\rm \Omega_{ta}$ with redshift and compare it to the prediction of the SCM for the Virgo simulations and for halos with $\rm \alpha_{3D} \geq 0.5$. With this cutoff, the offset between the points and solid lines is practically entirely eliminated.}
\label{Fig. 3.}
\end{figure}

The implementation of this cut is shown in Fig.~\ref{Fig. 3.}, where as in the upper panel of Fig.~\ref{Fig. 1.}, we plot the evolution $\rm \Omega_{ta}$ with redshift for halos from the Virgo simulations. When compared with Fig.~\ref{Fig. 1.}, it is clear that the offset has been essentially eliminated at all redshifts. Importantly, simulations with almost identical cosmologies (MDPL2 and Virgo $\rm \Lambda$CDM) but very different offset behaviors yield practically identical results after the implementation of this same single quality cut (see the blue and black points in Fig.~\ref{Fig. 3.}). We can thus conclude that the asphericity hypothesis provides a unified cosmology-independent explanation for the origin of the offset.   

With the offset eliminated, the properties of the $\rm \rho_{ta}(z)$ as a cosmology probe, as discussed in \cite{PavlidouEtal2020} using the SCM, now emerge in the simulation results as well: A measurement of $\rho_{ta}$ in a sample of sufficiently spherical halos at a single redshift (z=0) informs us of the value of the dark matter density in the universe. A few such measurements at higher z (up to a $\rm z=0.5$) are sufficient to distinguish between universes with the same dark matter content, but a different cosmological constant.

\section{Conclusions}
This work is  a follow-up to \citet{Korkidis_etal}, where we used N-body simulations for three different cosmologies to measure the turnaround density of cluster-sized halos in different redshifts, compared it to the predictions of the spherical-collapse model, and addressed the deviations between spherical collapse and simulations in different cosmologies. Specifically, we tested $\rm \Lambda CDM$, $\rm OCDM,$ and $\rm SCDM$ simulations, and analyzed snapshots of these at various $\rm z \leq 1$. 

We found that the halo dark matter density at the turnaround approximately agrees with the prediction of the spherical collapse model for all cosmologies. However, we detected an offset between $\rm \rho_{ta}$ in simulations and SCM that evolves (decreases) with redshift. We found that the offset is strongly anti-correlated with the degree of spherical symmetry in the distribution of neighbors. By applying quality cuts to the sample of simulated clusters for which we measured $\rm \rho_{ta}$, and in particular, by excluding halos whose distribution of massive neighbors was too aspherical, we eliminated (on average) this offset for all cosmologies and all redshifts. Because this selection criterion is independent of cosmology and redshift, its use restores the potential of $\rm \rho_{ta}$ as a potential cosmological observable.

\begin{acknowledgements}
We acknowledge support by the Hellenic Foundation for Research and Innovation under the “First Call for H.F.R.I. Research Projects to support Faculty members and Researchers and the procurement of high-cost research equipment grant”,  Project 1552 CIRCE (GK, VP); by the European Research Council under the European Union's Horizon 2020 research and innovation programme, grant agreement No. 771282 (KT); and by the  Foundation of Research and Technology - Hellas Synergy Grants Program (project MagMASim, VP, and project POLAR, KT).
The CosmoSim database used in this paper is a service by the Leibniz-Institute for Astrophysics Potsdam (AIP).
The MultiDark database was developed in cooperation with the Spanish MultiDark Consolider Project CSD2009-00064.
The authors gratefully acknowledge the Gauss Centre for Supercomputing e.V. (www.gauss-centre.eu) and the Partnership for Advanced Supercomputing in Europe (PRACE, www.prace-ri.eu) for funding the MultiDark simulation project by providing computing time on the GCS Supercomputer SuperMUC at Leibniz Supercomputing Centre (LRZ, www.lrz.de).
The Bolshoi simulations have been performed within the Bolshoi project of the University of California High-Performance AstroComputing Center (UC-HiPACC) and were run at the NASA Ames Research Center.
The simulations in this paper were carried out by the Virgo Supercomputing Consortium using computers based at Computing Centre of the Max-Planck Society in Garching and at the Edinburgh Parallel Computing Centre. The data are publicly available at \verb+www.mpa-garching.mpg.de/galform/virgo/int_sims+
Throughout this work we relied extensively on the PYTHON packages Numpy \citep{Numpy}, Scipy \citep{Scipy} and Matplotlib \citep{Matplotlib}. 
\end{acknowledgements}

%
%

\bibliographystyle{aa}
\bibliography{bibliography}

\begin{appendix}  
\section{Full correlations of turnaround mass and turnaround radius}
\label{appendix:a}
\begin{figure*}[htb!]
    \centering
    \includegraphics[width=0.7\textwidth,height=0.7\textheight]{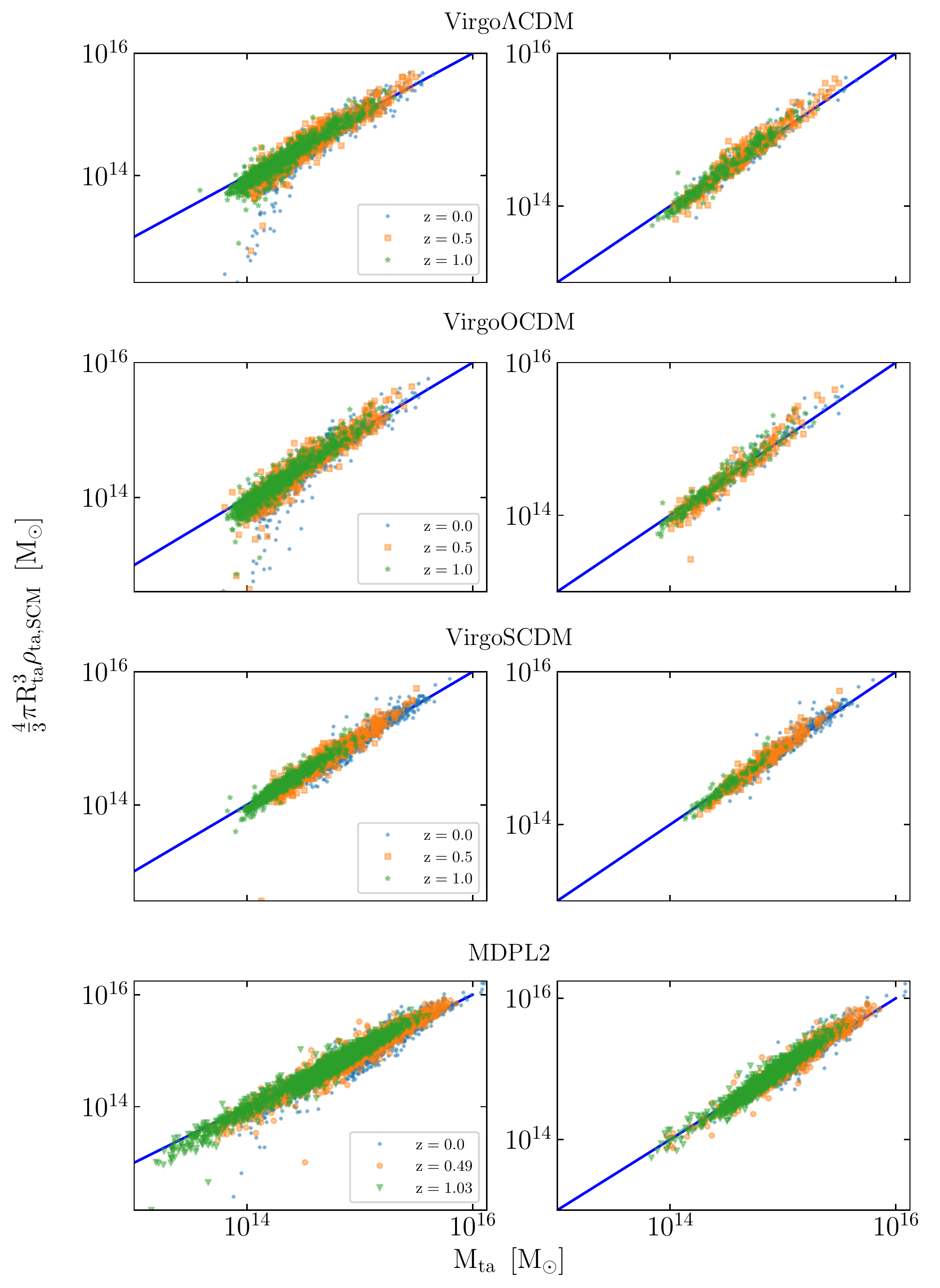}
\caption{Correlation of the turnaround mass and turnaround radius. Different colors and symbols represent different redshifts. Different rows depict different simulated cosmologies, as in the titles. In order to have the same scale for different redshifts, the third power of the turnaround radius was multiplied by the spherical-collapse--predicted turnaround density for the given cosmology and redshift. The right panel in each row shows the least-aspherical subsample of the structures ($\rm \alpha_{3D} \geq 0.6$; see \S \ref{subsection 3.2}). The blue line represents the $y=x$ line.}
\label{Fig. A.1.}
\end{figure*}

In \citet{Korkidis_etal}, the turnaround radius and mass were shown to scale as ($\rm R_{ta}^3 \sim M_{ta}$) so that a characteristic density on the turnaround scale of a structure can be defined. In this appendix, we verify that this remains true across redshifts in all simulations and cosmologies considered here. 
Figure \ref{Fig. A.1.} shows the quantity $4/3 \pi R_{\rm ta}^2 \rho_{\rm SCM}$ plotted against the turnaround mass $M_{\rm ta}$ for all simulations considered in this work, and for all the redshift snapshots we analyzed. $R_{\rm ta}$ is the kinematically defined turnaround radius, and $\rho_{\rm SCM}$ is the spherical-collapse model prediction for the turnaround density for the corresponding redshift and cosmology. For halos that behave exactly as the spherical collapse prediction, points should fall exactly on the $y=x$ line, which we overplot as the solid blue line. The left panel shows all halos in each snapshot, and the right panel only shows the least aspherical halos ($\alpha_{\rm 3D} \geq 0.6$, see \S \ref{subsection 3.2}).

\end{appendix}

\end{document}